\documentclass{aa}  

\usepackage{graphicx}
\usepackage{txfonts}
\usepackage{xcolor}
\usepackage[bookmarks=true,
  colorlinks,
  linkcolor=blue,
  urlcolor=blue,
  citecolor=blue,
  plainpages=false,
  pdfpagelabels,
  final,
  breaklinks=true]{hyperref}
\usepackage[capitalise]{cleveref}
\crefname{section}{Sect.}{Sects.}
\usepackage{array}
\usepackage{makecell}
\usepackage[normalem]{ulem}
\usepackage{booktabs}

\begin{document} 

   \title{Preparing for Rubin-LSST - Detecting Brightest Cluster Galaxies with Machine Learning in the LSST DP0.2 simulation}

   \author{A. Chu
          \inst{1,2}
          \and
          L. Doeser
          \inst{2}
          \and
          S. Ding
          \inst{3}
          \and
          J. Jasche
          \inst{2}
          }

    \institute{The Oskar Klein Centre, Department of Astronomy, Stockholm University, Albanova University Centre, 106 91 Stockholm, Sweden
    \and
    The Oskar Klein Centre, Department of Physics, Stockholm University, AlbaNova University Centre, 106 91 Stockholm, Sweden
    \and
    Sorbonne Universit\'e, CNRS, UMR 7095, Institut d'Astrophysique de Paris, 98bis Bd Arago, 75014, Paris, France
    }

    \date{Received XXXX; accepted XXXX}

    \abstract{The future Rubin Legacy Survey of Space and Time (LSST) is expected to deliver its first data release in the current of 2025. The upcoming survey will provide us with images of galaxy clusters in the optical to the near-infrared, with unrivalled coverage, depth and uniformity. The study of galaxy clusters informs us on the effect of environmental processes on galactic formation, which directly translates onto the formation of the brightest cluster galaxy (BCG). These massive galaxies present traces of the whole merger history of their host clusters, which can be in the shape of intra-cluster light (ICL) that surrounds them, tidal streams, or simply by the accumulated stellar mass that has been acquired over the past 10 billion years as they have cannibalized other galaxies in their surroundings.
    In an era where new data is being generated faster than humans can deal with, new methods involving machine learning have been emerging more and more in the most recent years. In the aim of preparing for the future LSST data release which will allow the observations of more than 20000 clusters and BCGs, we present in this paper different methods based on machine learning to detect these BCGs on LSST-like optical images. This study is done by making use of the simulated LSST Data Preview images. We find that the use of machine learning allows to accurately identify the BCG in up to 95\% of clusters in our sample. Compared to more conventional red sequence extraction methods, the use of machine learning appears to be faster, more efficient and consistent, and does not require much, if any, pre-processing. }


    \titlerunning{Machine Learning for BCG detection}
    
    \maketitle

%

\section{Introduction}

Galaxy interactions and mergers play a major role in galaxy formation and evolution. These events are believed to be the biggest contributing factor to the mass assembly of galaxies and to the morphological evolution of galaxies through the Hubble sequence \citep{SpitzerBaade1951ApJ...113..413S, Biermann1975A&A....41..441B, Hickson1977ApJ...213..323H, Marchant1977ApJ...215....1M, Roos1979A&A....76...75R, Dressler1984ARA&A..22..185D, Querejeta2015A&A...579L...2Q}. Although only a minority in the galaxy population, early-type galaxies (mostly old red elliptical galaxies with little to no gas and quenched star formation) dominate in high-density environments such as galaxy clusters, whereas late-type galaxies (mostly young blue spiral galaxies with a lot of gas and high star formation rates) dominate mostly in the field \citep{Hubble1936rene.book.....H, Dressler1980ApJ...236..351D, Rood1972ApJ...175..627R, Oemler1974ApJ...194....1O}. 

This morphological segregation originates from the higher galactic encounter likelihood in clusters compared to the field. 
Clusters are environmentally dense systems that contain hundreds to thousands of galaxies concentrated in a very small volume of the Universe, with sizes averaging just one or two Mpc in diameter. Because of the high number density of galaxies, galaxies in clusters are much more likely to experience mergers, collisions, or other gravitational interactions with other galaxies in their lifetimes than any other galaxies. It is now established that the merger of two spiral galaxies of similar masses gives birth to a massive elliptical galaxy \citep{Toomre1977egsp.conf..401T, sawala2023distinctdistributionsellipticaldisk}. The gas in these galaxies gets consumed to fuel a new starburst phase, or/and gets ejected in the process, leaving the resulting galaxy with a limited amount of gas. This explains the dominance of elliptical galaxies in galaxy clusters or groups. As a result, clusters are usually identified by an overdensity of red elliptical galaxies with similar colours, forming a sequence in a colour-magnitude diagram called the red sequence \citep{Baum1959PASP...71..106B, deVaucouleurs1961ApJS....5..233D}.

In the context of the large-scale structures, galaxy clusters are located in the nodes of the cosmic web, at the intersection of cosmic filaments. The growth of clusters is fueled by infalling matter from these filaments. This infalling matter consists of gas or small galaxies. As the cluster forms, and as more and more matter falls to the central region of the cluster, at the bottom of the cluster gravitational potential well, a supermassive galaxy forms. This central galaxy is perfectly located to receive and accrete all this matter and merges with smaller galaxies through time which contributes to its mass assembly. It often ends up becoming the brightest and most massive galaxy in the cluster, and as such, is referred to as the brightest cluster galaxy (BCG). 

In a hierarchical evolution scenario where small clumps of matter assemble to form bigger and bigger entities over time, BCGs are the final products of hierarchical evolution at the galactic scale. BCGs are the results of the merging histories of their host clusters and have properties closely linked to those of their host clusters \citep{lauer2014brightest, Chu+21, Chu+22, Durret_2019, west2017ten}. They can thus give us insights on how and when the clusters were formed, on the co-evolution of BCGs and their host clusters \citep{ sohn2020ApJ...891..129S,Sohn2022ApJ...931...31S, GoldenMarx2022ApJ...928...28G}, and help us better understand how the environment they are in can impact their formation, i.e, to better constrain the role of galactic mergers on galaxy evolution \citep{SpitzerBaade1951ApJ...113..413S}. They also constitute the most massive galaxies in the Universe, making them excellent probes for testing cosmological models \citep[see for example][]{Stopyra_2021}. 

Although very good proxies to study galaxy formation, the literature lacks statistically significant studies on these objects, not allowing us to properly constrain the properties of these massive galaxies and their formation histories. As a result, it is still unclear if BCGs are still evolving or growing today. Some authors find a significant growth in size since the last 8 Gyrs \citep{Ascaso_2010, Bernardi_2009, Nelson2002ApJ...567..144N, Yang2024MNRAS.531.4006Y}. Others, such as \citet{Stott_2011}, however, do not find any evidence of such a growth. In \citet{Chu+21} and \citet{Chu+22}, they find that while the physical properties of BCGs may not have evolved much since $z = 1.8$, their structural properties may have changed in more recent times with the formation of a faint and diffuse stellar halo of the BCG that is not associated with intracluster light. This halo is believed to be the product of more recent mergers with smaller satellites. This result is in favour of an inside-out growth scenario suggested by many authors \citep{bai2014inside, lauer2014brightest, 10.1093/mnras/stz2706, DeMaio2020MNRAS.491.3751D}. Constraining the epoch of formation of this halo, as well as the rate at which it formed, is however limited by the lack of big and deep surveys that do not allow us to detect this dual structure in BCGs.

This problem may be no more thanks to the future Rubin Legacy Survey of Space and Time (LSST). The Rubin-LSST survey, with the telescope expected to see its first light in the course of 2025, will provide images of the whole southern hemisphere with unprecedented depth in six different bandpasses, from the optical to the near-infrared, $ugrizy$. At the end of its 10-year-long mission, Rubin-LSST is expected to detect about 20000 galaxy clusters with a surface brightness limit in the $r$ band reaching 30.5 mag/arcsec$^{2}$ \citep[See][for a full description of the LSST survey]{IvezicLSST2019ApJ...873..111I}.
On the contrary of most wide-field surveys, such as the Sloan Digital Sky Survey (SDSS), the Hyper Suprime Cam - Subaru Strategic Program (HSC-SSP) or the Dark Energy Survey (DES), which prioritize volume over depth, Rubin-LSST will combine both. Its large footprint will enable us to increase statistical studies on clusters and BCGs, and its depth will allow us not only to better measure the photometric and structural properties of these objects, but also to better constrain their evolution with time as we will observe BCGs not only in the local Universe, but also up to the epoch of their formations in proto-clusters ($z > 2$).

To keep up with the fast cadence of Rubin-LSST and the significant amount of data that will be generated, new methods to reduce and analyze the data are necessary. Indeed, we expect no less than 15 TB of data to be acquired each night, and the whole southern sky to be surveyed in only three nights in two photometric bands. Methods to detect galaxy clusters as well as BCGs do exist in the literature. Those are mostly based on detecting the red sequence based on photometric measurements of their colours \citep[e.g RedMapper, see][]{Rykoff_2014}. However, these algorithms tend to be very limited in redshift (up to $z \sim$ 0.7), as in order to extract the red sequence, filters in the infrared would be necessary to frame the 4000\AA~break. Also, BCG detection algorithms, in particular, tend to contain a not so insignificant amount of false detections and require a visual inspection \citep[][for example has a 70\% success rate]{Chu+22}. With the quantity of data that will have to be processed with Rubin-LSST, this task is doomed to be futile and unreasonable. 

In more recent years, the use of machine learning in multiple domains has been on the rise. Machine learning allows us not only to analyze a large volume of data in a reasonable timeframe but also to identify some patterns that may have otherwise been missed by the human eye. The use of machine learning in astrophysics and cosmology domains has already been seen in the literature \citep[][]{ntampaka2021rolemachinelearningdecade, Moriwaki2023RPPh...86g6901M,HuertasCompany2023PASA...40....1H, baron2019machinelearningastronomypractical}. In particular, recent relevant works include the detection of ICL and BCGs in simulated clusters \citep{Marini2022MNRAS.514.3082M}, on Hyper Suprime-Cam images \citep{canepa2025measuringintraclusterlightfraction}, combining both simulations and observations from the SDSS \citet{2025arXiv250200104J}, or the detection of proto-clusters in simulation and observational data \citep{Takeda_2024}.

In this paper, we present multiple machine learning-based algorithms in order to detect BCGs on Rubin-LSST images and show how machine learning is proved to be an excellent and more efficient tool for BCGs detection compared to more traditional methods. In \cref{section:datadp02}, we present the data used in this study. We explain the procedure used to prepare and increase the data sample for machine learning models in \cref{section:dataprocess}. In \cref{section:detection_ml}, we detail different machine learning algorithms which aim to detect BCGs on images. Then, in \Cref{section:perf}, we compare the performance of our machine learning methods with more traditional methods. In \cref{section:discussion_conclusion}, we discuss our results and present our conclusions.


\section{LSST Data Preview DP0.2}
\label{section:datadp02}

In preparation for the first Rubin-LSST data release, we make use of the LSST Data Preview DP0.2 to test our algorithms. We here describe briefly the procedure detailed on the LSST DP0.2 website\footnote{\href{https://dp0-2.lsst.io}{https://dp0-2.lsst.io}}, used to generate the LSST DP0.2. 

The LSST DP0.2 is a 300 deg$^{2}$ simulated survey based on images generated by the Dark Energy Science Collaboration (DESC) in the context of the second data challenge (DC2) \citep[see the description paper][]{Abolfathi_2021,lsstdarkenergysciencecollaboration2022descdc2datarelease}. The DC2 is based first on a large cosmological $N$-body simulation, the Outer Rim simulation \citep{2019ApJS..245...16H}, which is used to generate an extragalactic catalogue, the cosmoDC2 catalogue \citep{2019ApJS..245...26K}. The cosmoDC2 catalogue is then fed to the LSST software \texttt{CatSim}\footnote{\href{https://www.lsst.org/scientists/simulations/catsim}{https://www.lsst.org/scientists/simulations/catsim}} to generate instance catalogues, i.e, extragalactic catalogues that take into account observational constraints depending on the location of a source in the sky and the cadence of the telescope over the 10-year-long mission of Rubin-LSST. As such, photometry will include extinction due to galactic dust, astrometric shifts due to the motions of the Earth, and also uncertainties estimates of the sources' luminosities and fluxes. \texttt{CatSim} also adds other galaxy features and time variability not present in the original cosmoDC2 catalogue. These instance catalogues are then passed through the ImSim image simulation tool, which renders these catalogues into LSST-like images. These steps thus bring the images to LSST-resolution, with LSST-like noise. LSST DP0.2 images are simulated images corresponding to 5 years of LSST observations obtained with a simulated LSST cadence.

In this study, we make use of the cosmoDC2 catalogue in order to retrieve the positions of the BCGs and their clusters in the LSST DP0.2 survey. We retrieve all halos in the cosmoDC2 catalogue of a mass M $\geq$ 10$^{14}$ M$_{\odot}$. 
The extracted cluster sample consists of 8072 clusters. 

Deep coadded images of the clusters in all 6 $ugrizy$ filters are then produced using the LSST pipelines \citep{2018PASJ...70S...5B, 2019ASPC..523..521B, 2022SPIE12189E..11J}, available on the Rubin Science Platform (RSP) \citep[Jurić et al. 2019\footnote{\href{https://lse-319.lsst.io}{https://lse-319.lsst.io}}, Dubois-Felsmann et al. 2019\footnote{\href{https://ldm-542.lsst.io}{https://ldm-542.lsst.io}}][]{2024ASPC..535..227O}. All images are centred on the BCG coordinates from the cosmoDC2 catalogue, and have a fixed physical size of 1.5 Mpc. Because of out-of-memory issues while creating the deep-coadded images, 25 clusters could not be added to our final sample. Only considering clusters found in the footprint of the DP0.2 survey, the final sample contains 6348 clusters between redshifts 0 $\leq$ z $\leq$ 3.6. The coordinates of the BCGs in the clusters are given in the cosmoDC2 catalogue.


\section{Data processing}
\label{section:dataprocess}

Some key parameters in machine learning to efficiently train a model are the data samples provided, its uniformity, and its size. Big samples are needed to increase the accuracy of a model. Insufficient data may lead to the neural network overfitting, i.e., learning to adapt to the training data too closely, not allowing it to return accurate predictions to a new set of data that is not used for the training. In this case, the neural network is unable to converge to a generalised solution of the problem due to the under-sampled training set. In instance, \citet{2015arXiv151106348C} discusses how to estimate how much data is needed to obtain high accuracy in deep learning problems.

In an effort to increase our training dataset, we apply several transformations and procedures to our initial image sample. These transformations include: centre shifts, implementation of redshift uncertainties, as well as image rotations.

We first add uncertainties to the redshift of the cluster, to mimic photometric redshifts as measured by the LSST pipelines. The new redshift has an uncertainty randomly generated from a uniform distribution between $z - 0.05 \leq z \leq z + 0.05$, which is slightly bigger than the expected photometric uncertainties of LSST estimated at about 0.02. We then generate a centre shift randomly chosen up to 300 kpc in physical distance from the centre of the cluster as defined in the cosmoDC2 catalogue. This physical distance is calculated using the angular diameter obtained using the newly generated redshift. The image is then cropped to a physical distance of 1.5 Mpc in size. A random angle is also generated between 0 and 180 degrees to rotate the final image. Finally, the image is resized so it will have final dimensions of $512 \times 512$ pixels. These transformations are applied to all $ugrizy$ bandpasses, and we also retrieve the new coordinates in pixels of the BCG on these images. This procedure is repeated ten times for each cluster, thus increasing our cluster sample by a factor of ten to a total of 63480 clusters.

Because proto-clusters ($z > 2.0$) are much different systems than clusters, with extents reaching several Mpc, up to 20 Mpc as these systems are still actively forming, and because their BCGs are in the process of merging and growing and thus have significantly different properties than cluster BCGs, we have decided to only select for this study clusters up to a redshift $z \leq$ 2.0. 

The resulting full sample of 63140 clusters is divided into three sets: a training sample, a validation sample, and an evaluation sample of respectively 55778 ($88.3\%$), 2000 ($3.2\%$) and 5362 ($8.5\%$) clusters. These three samples have comparable redshift distributions.


\section{Detection of BCGs with machine learning}
\label{section:detection_ml}

Rubin-LSST will provide us with images in six filter bands, $ugrizy$, of the $18\,0000\,\rm{deg}^2$ surface area in the southern hemisphere. We can expect to get new observations of more than 20000 galaxy clusters and BCGs that will be studied in a uniform way. 

Detecting BCGs on images is not as straightforward as it may seem. Spectroscopy, especially for a field big enough to cover a cluster, is hard to obtain, so the development of efficient algorithms that can be applied to photometric surveys is becoming more and more important. Images contain various sources and artefacts that can make the task difficult (stars, diffraction spikes, foreground and background galaxies, etc.). Visually, galaxy clusters are easily identifiable as overdensities of red elliptical galaxies in the sky. This particular characteristic of clusters is usually used to detect clusters in photometric surveys via the red sequence \citep{Baum1959PASP...71..106B, deVaucouleurs1961ApJS....5..233D}. The red sequence is drawn on a colour-magnitude diagram by all the red elliptical galaxies of the cluster which present a very small dispersion in their colours. It indicates which galaxies are likely to be part of the cluster based on their colours. 

Generally, these red sequence-based algorithms are quite efficient in detecting isolated clusters at lower redshifts (no superclusters or superposition of two clusters at two different redshifts), but are less efficient for more complex cases and at higher redshifts where photometric measurements (magnitudes in occurrence) become less reliable as the red sequence is less defined. Other sources in the background or foreground can also have colours similar to those of red sequence galaxies at a certain redshift because of unfortunate well-placed emission lines which can boost the magnitude in a given filter. On top of detecting the cluster, detecting the BCG adds another layer of complexity in the process. The BCGs, which are commonly believed to be located in the central region of the cluster, may not be in the centre but displaced because of a recent cluster or galactic merger \citep{Patel_2006, Hashimoto_2014, De_Propris_2020, Chu+21}. This BCG may also not appear to be much different from the other galaxies in the cluster in terms of luminosity or size, which makes its identification all the more difficult.

Because of the size of the upcoming survey, identifying each BCG visually and individually on images is not possible. Many algorithms have thus been developed in this aim \citep[see][among others]{Rykoff_2014,Chu+21,Chu+22}, but those are quite limited as they are still heavily dependent on photometric and derived properties measured on images. We aim to build one of, or even, the largest and purest catalogue of BCGs to date, in order to better model these galaxies and better understand their formation. The big sky area covered by the Rubin-LSST survey will enable us to construct a catalogue of significant size, and we use a new machine learning-based algorithm in order to detect BCGs on images with high efficiency, returning a large catalogue with high purity.

\begin{figure*}
\centering
\includegraphics[width=\hsize]{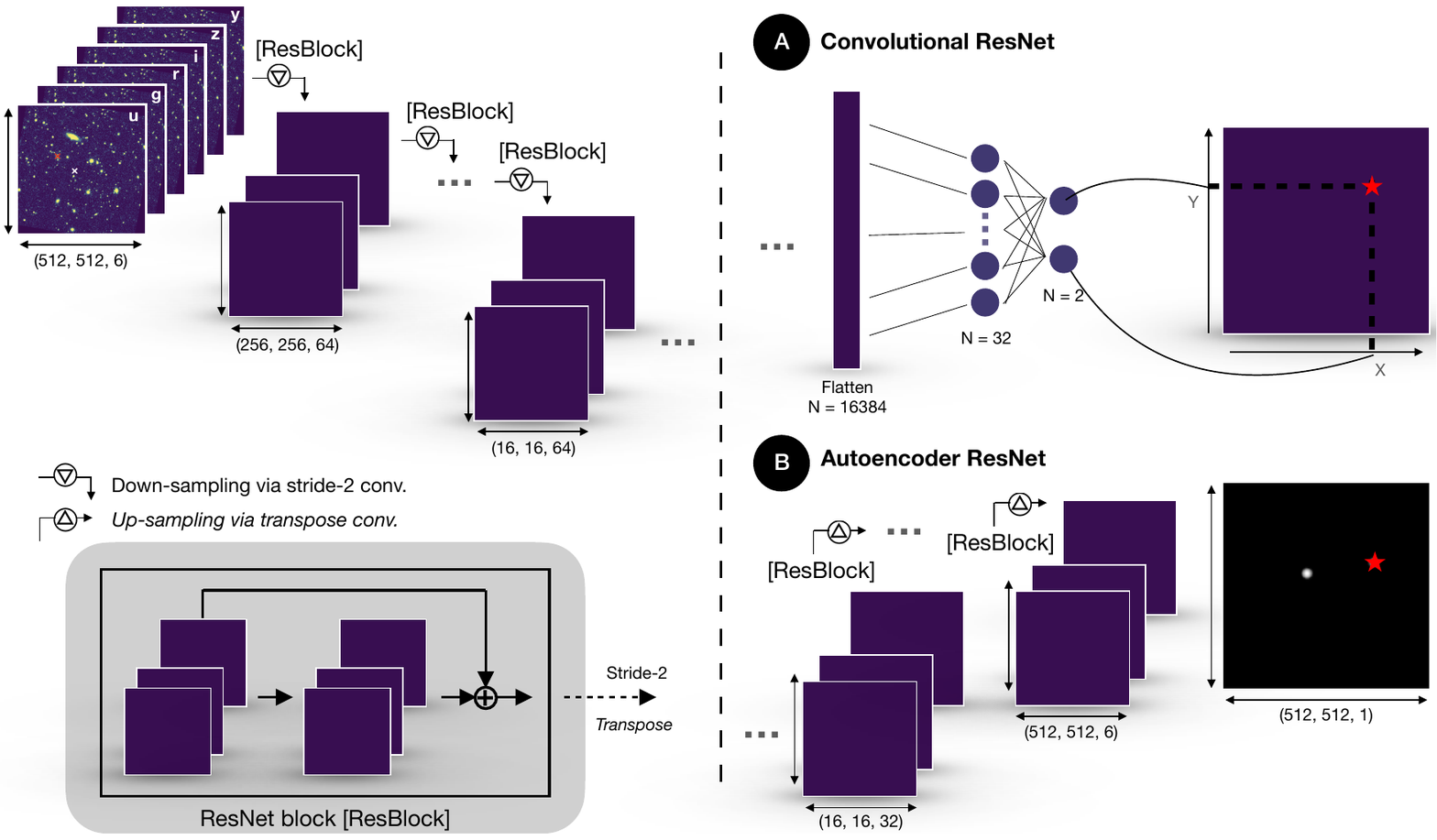}
\caption{Architecture of the two neural networks used in this paper: the convolutional ResNet (A-branch) presented in \Cref{subsection:cnn_resnet} and the autoencoder ResNet (B-branch) presented in \Cref{subsection:cnn_autoencoder}. Both networks are based on a residual deep learning architecture detailed in the legend on the bottom left-hand side. The architecture of both networks is similar in the first half and differs in the second half as indicated by the two different branches.}
\label{fig:ml_arch}
\end{figure*}

The codes used in this paper and detailed hereafter can be made available upon reasonable request. The architecture of the two networks presented in this paper is illustrated in \cref{fig:ml_arch}.

\subsection{Convolutional residual neural network}
\label{subsection:cnn_resnet}

In the first part, we use a supervised convolutional neural network to return the (x, y) pixel coordinates of the BCG in images. The neural network architecture consists of a 5-layer convolutional residual neural network (ResNet) \citep{he2015deepresiduallearningimage}, topped with a dense layer. Each residual block is made of two 2D-convolutional (Conv2D) layers, each Conv2D layer is followed by a dropout layer. The first Conv2D+DropOut layer feeds into the second. The output of the first Conv2D+DropOut layer is concatenated with the output from the second one. The output of this concatenation then goes through a pooling layer. The Conv2D layers have a "relu" \citep{2018arXiv180308375A} activation function. The pooling is done using another 2D-convolutional layer with stride 2. Although this slightly increases the overall network parameters, using convolutional layers with stride larger than 1 effectively means learning a pooling kernel, which has been demonstrated to increase the expressibility of CNNs \citep{he2015deepresiduallearningimage, milletari2016vnetfullyconvolutionalneural}. The final part of the network is composed of a flattening layer, a dense layer, followed by a dropout layer and finally, a dense layer with 2 neurons to output the (x, y) BCG coordinates. We use a learning rate of $lr$ = 0.0001, a mean squared error loss function and the Adam optimiser \citep{kingma2017adammethodstochasticoptimization}.

The use of a ResNet \citep{he2015deepresiduallearningimage} allows us to better deal with vanishing gradients which prevent the accuracy of deep neural networks to improve with complexity. In a traditional neural network, one layer feeds directly into the next and so on. In that case, in a $N$-layer network with the initial input being denoted $x$, each layer gets as an input $f_{n}(x)$, with 1 $\leq n \leq N$, such as the network tries to directly learn the function that will produce the desired output $H(x)$. With residual blocks, we incorporate skipped connections: one layer feeds into another layer, but instead of only considering the output of this second layer, we keep the identity of the first one as well. In that case, one layer gets as an input $f(x) + x$, and the identity is preserved through the following layers as well. The output of each residual block is a global function of $x$ and the identity $x$, $f_{n}(x) + x$. As a result, instead of trying to directly match the output such as $F(x) = H(x)$ in a traditional network; in a ResNet, we calculate a global residual function instead as $H(x) = F(x) + x$ and $F(x) = H(x) - x$. The use of a ResNet with skipped connections enables to better retain information present and learnt in the first layers of the neural network, which in a traditional network may end up getting lost due to small weights in a layer. It also enables one to skip entirely one layer if unnecessary, which, in a traditional network is not feasible as all layers depend on the output of the previous ones. 

The ResNet takes as inputs images in the cluster in all six $ugrizy$ bandpasses, of dimensions 512$\times$512 pixels, and returns as outputs the pixel coordinates (x, y) of the BCG. 

\begin{table*}[htbp]
\centering
\caption{Table summarizing the results from the three different detection methods used in this paper: the convolutional ResNet detailed in \Cref{subsection:cnn_resnet}, the autoencoder ResNet described in \Cref{subsection:cnn_autoencoder}, and the red sequence based algorithm used as benchmark and discussed further in \Cref{subsection:detection_rs}. The columns are, from left to right: the method used, the input and the output of each method, the number of parameters in the neural network, the accuracy defined as the number of good detections over the total number of clusters in the evaluation sample, and finally the pros and cons of each method.}
\label{tab:detect_summ}
\begin{tabular}{| m{2cm} || m{2cm} | m{1.7cm} | m{1.2cm} | m{0.7cm} | m{3.9cm} | m{3.9cm} |}
\hline
& & & & & & \\
Method & Input & Output & Param. & Acc. & Pros & Cons \\
& & & & & &\\
\hline     
\hline

& & & & & & \\
 Convolutional ResNet & 6 filters $ugrizy$ 512$\times$512 images & $(x,y)$ pixel coordinates & 830,882 & 81\% & \begin{itemize}
     \item Good accuracy
     \item Straightforward
     \item No redshift dependencies
 \end{itemize} & \begin{itemize}
     \item No flag for bad detections
     \item Gets confused by the presence of multiple candidates
     \item Training is computationally expensive
     \item Difficulties to detect sources with low-SNR and deblending issues
     \end{itemize} \\
 \hline
 Autoencoder ResNet & 6 filters $ugrizy$ 512$\times$512 images & Probability map & 531,719 & 95\% & \begin{itemize}
     \item Very good accuracy and resolution
     \item Provides a confidence level for each detection
     \item Can detect multiple BCG candidates
     \item No redshift dependencies
 \end{itemize} & \begin{itemize}
     \item Difficulties to detect sources with low-SNR and deblending issues
 \end{itemize} \\ 
 \hline
 Red Sequence algorithm & Photometric catalogues in at least 2 filters & BCG ID in the catalogue & & 70\% & \begin{itemize}
     \item Physically motivated
 \end{itemize} & \begin{itemize}
     \item Lower accuracy
     \item Requires pre-processing of photometric catalogues
     \item Difficulties to detect sources with low-SNR and deblending issues
     \item Photometry less reliable at high redshifts
     \item Need redder filters to contain the 4000\AA~at higher redshifts
     \end{itemize} \\ 
 \hline
\end{tabular}
\end{table*}

\begin{figure*}[h]
  \centering
  \begin{tabular}{@{}c@{}}
    \includegraphics[width=.497\linewidth]{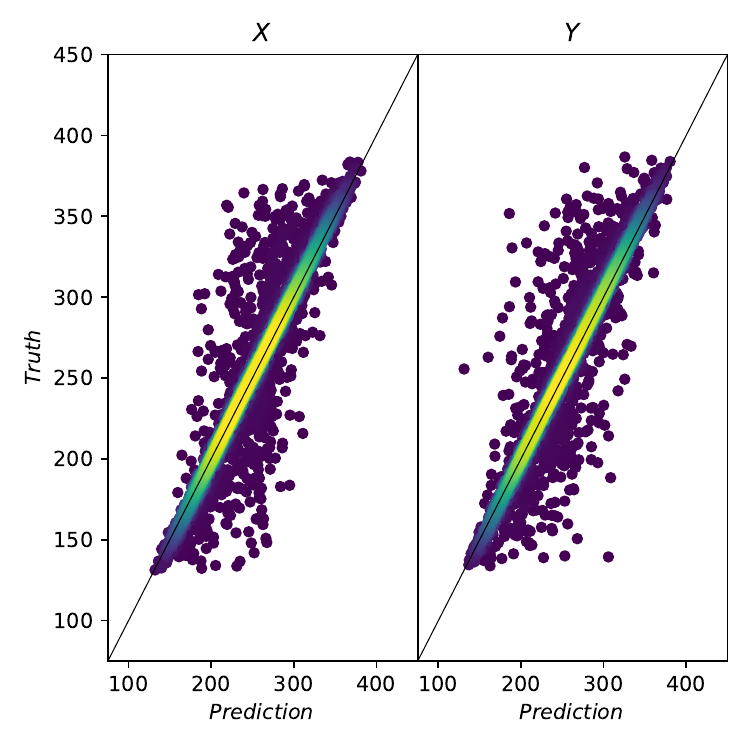}
  \end{tabular}
  \begin{tabular}{@{}c@{}}
    \includegraphics[width=.497\linewidth]{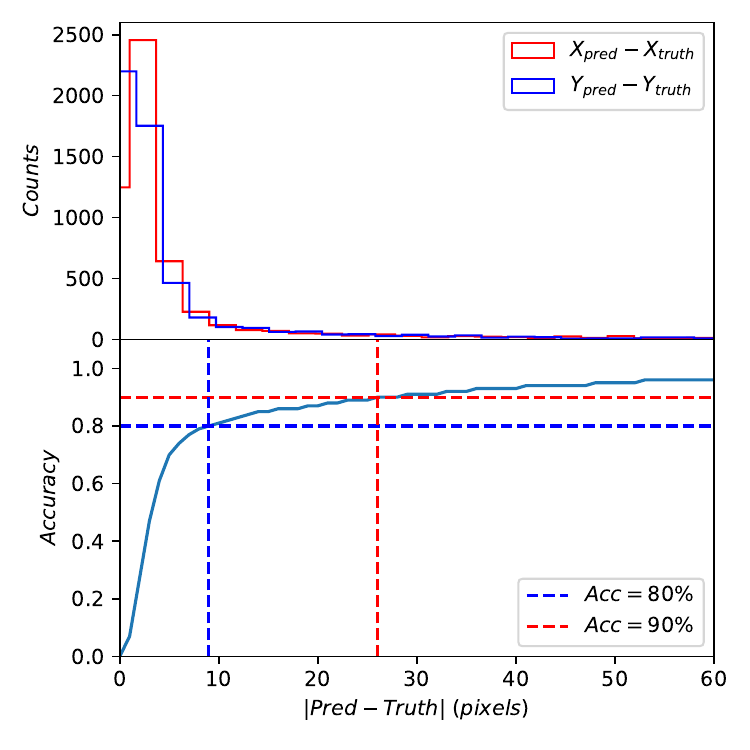}
  \end{tabular}
  \caption{Accuracy of the convolutional ResNet described in \Cref{subsection:cnn_resnet}. \textit{Left:} $(x, y)$ predicted coordinates of the BCGs as a function of their true $(x, y)$ coordinates as given in the cosmoDC2 simulation. The figure shown here is obtained for the BCGs in the evaluation sample. The color bar represents the number density of points.  \textit{Right}: (Top panel) Histogram of the difference between the predicted and true (x, y) coordinates of the BCGs in the evaluation sample as detected by the ResNet. The red and blue histograms represent the $x$ and $y$ coordinates respectively. (Bottom panel) Accuracy of the convolutional ResNet depending on the defined resolution considered for a well-detected BCG. The blue and red lines correspond to an accuracy of 80\% and 90\% respectively.}
    \label{fig:resnet_predtruth_histdiff_acc}
\end{figure*}

\subsection{Convolutional residual autoencoder}
\label{subsection:cnn_autoencoder}

In a second part, in an effort to increase the accuracy of the convolutional ResNet and using as a baseline the architecture described in \Cref{subsection:cnn_resnet}, we develop a residual autoencoder that, instead of outputting (x, y) coordinates, returns a probability map with the same inputs. We create two different types of probability maps and compare the results of both in the following section.

First, the probability maps are simply created by drawing a small circular source with a radius of 10 pixels at the position of the BCG, and convoluting the profile with a Gaussian, with a kernel size of 150 pixels and a standard deviation $\sigma$ = 4. The parameters of the Gaussian were chosen so it would be big enough to cover most of the center of the BCG profile, and so that the probability in a radius of 5 pixels would be of about 50\%, 5 pixels being the desired resolution. 

Then, for comparison, and in an attempt to get even more accurate outputs, the probability maps are obtained by using Gaussian convoluted segmentation maps. The segmentation maps are obtained from the detection code \texttt{SourcExtractor} \citep{BertinArnouts96}. Whereas in the first method, the neural network learns to mainly recognise the central region of the BCG, here it will be trained using the whole profile and morphology of the BCG. These segmentation maps are obtained only in the $z$ bandpass, in the reddest bandpass, as it is the one with the highest signal-to-noise ratio. We thus assume the same segmentation map for all 6-bandpasses. They flag the pixels associated with the BCG as ones and the rest as zeros. Then, we convert these segmentation maps into probability maps by convolving the segmentation maps with a Gaussian profile, as described in the first method. For clusters in which the BCG could not be detected because of deblending issues, the segmentation map is created by drawing a smaller circular source with a radius of 5 pixels, and convoluting the profile with a Gaussian. We choose to still include these clusters in the training to infer if the neural network actually manages to outperform usual detection algorithms which tend to miss these sources.

We here use a convolutional residual autoencoder composed of an encoder which will extract the main features from the input, and a decoder which will reconstruct the same dimension input using the features extracted by the encoder. The encoder architecture is the same as the ResNet described in \Cref{subsection:cnn_resnet}. The decoder that follows the encoder differs only by the replacement of Pooling layers into Transpose layers to up-sample instead of down-sample. The final layer of the autoencoder is a Transpose layer that will return 6 probability maps in all 6-$ugrizy$ filterbands. We use one last Conv2D layer to compress this output into a single probability map on which the BCG will be detected. The loss function used here is a binary cross-entropy loss function.

The central coordinates of the BCG will be associated with the pixel with the highest probability.

\section{Performance of the neural networks}
\label{section:perf}

We here discuss the performance of both neural networks described in \Cref{section:detection_ml}. We apply the trained neural networks on the evaluation data of 5362 galaxy clusters. In both cases, we define an accurately detected BCG if the coordinates of the predicted BCGs are within 10 pixels of the true coordinates of the BCG. The accuracy is thus defined as the number of BCGs whose $(x, y)$ coordinates are less than 10 pixels from the true position of the BCG, over the total number of BCGs in the evaluation sample. The results are summarized in \Cref{tab:detect_summ}.

\subsection{Convolutional residual neural network}
\label{subsection:perf_resnet}

We show on the left panel of \Cref{fig:resnet_predtruth_histdiff_acc} the comparison between the $(x, y)$ coordinates of the BCGs in the evaluation sample as predicted by the convolutional ResNet described in \Cref{subsection:cnn_resnet} and the true coordinates. We find a good correlation between the predicted and true coordinates, as most points are aligned along the main axis. Both $x$ and $y$ coordinates have a very strong correlation coefficient between the predicted and true coordinates of respectively $r_{x}$ = 0.95 and $r_{y}$ = 0.96. 

The distribution of the difference between the $(x, y)$ predicted and true coordinates is shown on the right top panel of \cref{fig:resnet_predtruth_histdiff_acc}. The distribution is a Gaussian with a mean and standard deviation of $\left(\mu_{x} = -1.51,\sigma_{x} = 16.22\right)$ and $\left(\mu_{y} = -1.53, \sigma_{y} = 15.56\right)$ for the $x$ and $y$ coordinates respectively.

The bottom right panel of \cref{fig:resnet_predtruth_histdiff_acc} shows the accuracy of the ResNet depending on the resolution defined to consider a BCG as accurately-detected. We find that, for the defined resolution of 10 pixels, 81\% of the BCGs in the evaluation sample have been accurately detected. In order to obtain an accuracy of at least 90\%, we would need to increase the resolution to up to 26 pixels. However, this limit is too big as the BCG halo may extend up to 20 pixels on our images, so at these distances, the predicted position would fall outside of these borders.

\begin{figure}
\centering
\includegraphics[width=\hsize]{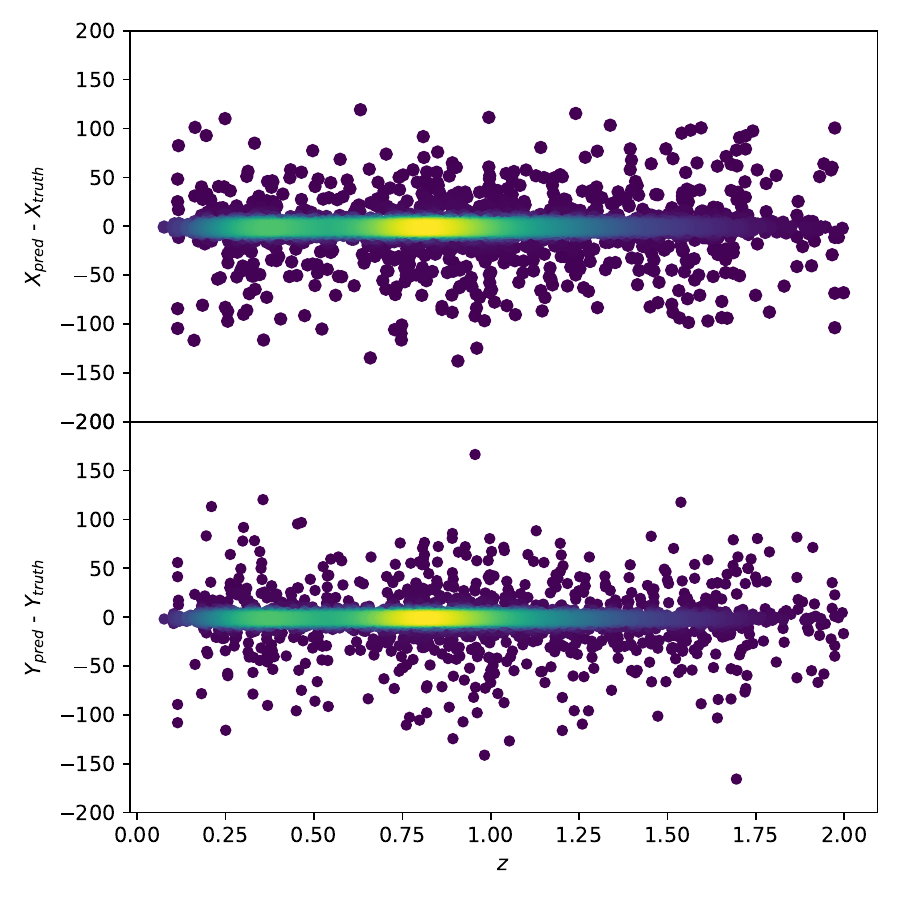}
\caption{Difference between the $(x, y)$ predicted and true coordinates of the BCGs in the evaluation sample, using the convolutional ResNet, and the clusters' redshifts. The color bar represents the number density of points.}
\label{fig:resnet_zdiff}
\end{figure}

In \cref{fig:resnet_zdiff}, we verify if the outliers with big differences are associated with the clusters at higher redshift. Indeed, at high-$z$, galaxies are less resolved, the signal-to-noise ratio is lower, we expect photometric measurements to be less precise and also the red sequence to not be as defined because of not red-enough bandpasses to constrain the 4000\AA~break. All these reasons may cause the neural network to not train as efficiently at these distances. Indeed, with skewed photometry, distinguishing between field galaxies and cluster galaxies in the red sequence would be more complicated, and detecting overdensities of galaxies would also appear to be a harder task as fainter objects may not be detected. However, we find that outliers are found at all redshifts, without any distinction between low-$z$ and high-$z$ clusters.

\begin{figure*}[t]
\centering
\includegraphics[width=\hsize]{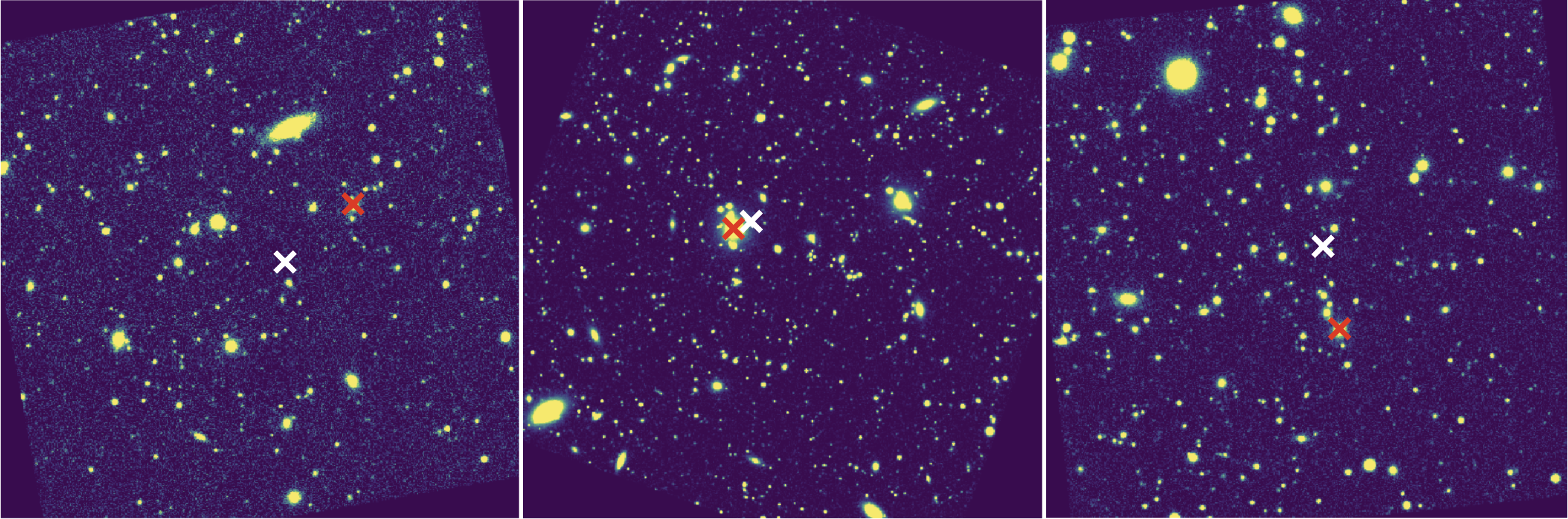}
\caption{Example of three clusters in the LSST-DP0.2 simulation in which the predicted BCG indicated by the white cross is offset from the true BCG from the cosmoDC2 catalogue indicated by the red cross.}
\label{fig:resnet_detecthaloex}
\end{figure*}

A visual inspection of the outliers shows that clusters that present the biggest offsets tend to be mostly clusters with more complex structures where two or more galaxies appear to be comparable to the BCG as defined in the cosmoDC2 catalogue. Indeed, in not so few cases, BCGs in clusters may not be much different from the second or third brightest galaxies, making the distinction with the actual BCG not so straightforward. Actually, depending on the definition of a BCG, the identification of a BCG may differ. In the cosmoDC2 catalogue, the BCG is defined as the brightest central galaxy. However, some authors would argue that the BCG should be the brightest galaxy of the cluster, independently of its distance from the centre of the cluster. In such clusters, where several galaxies appear to be equally good BCG candidates, we find that the ResNet returns a BCG position that is located in between of these candidates. Such examples are given on \cref{fig:resnet_detecthaloex} where the detected BCG (red cross) appears offset from the true BCG (white cross) because of the presence of a comparable galaxy in the field of the cluster. We note that, in such complex cases, the detection of the true BCG or the second BCG candidate would both be acceptable as both galaxies appear to be very similar in sizes, luminosities, and location in the cluster relative to the galaxy density distribution. However, as can be seen on \cref{fig:resnet_detecthaloex}, the network returns an in-between position that does not correspond to any object. Although this neural network appears to be relatively efficient, with an accuracy of 81\%, it does not allow us to automatically identify the bad detections.

\subsection{Convolutional residual autoencoder}
\label{subsection:perf_autoencoder}

\begin{figure*}[h]
  \centering
  \begin{tabular}{@{}c@{}}
    \includegraphics[width=.497\linewidth]{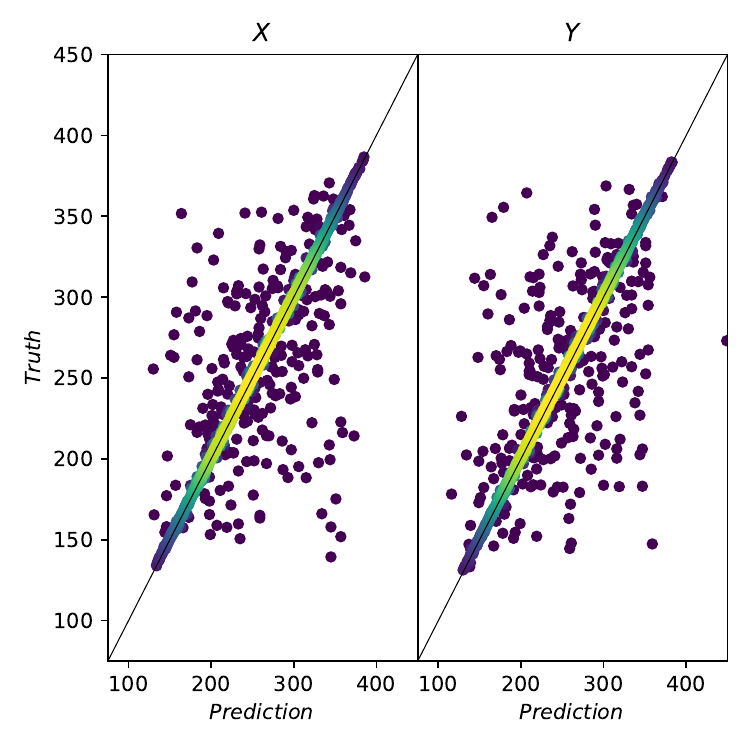}
  \end{tabular}
  \begin{tabular}{@{}c@{}}
    \includegraphics[width=.497\linewidth]{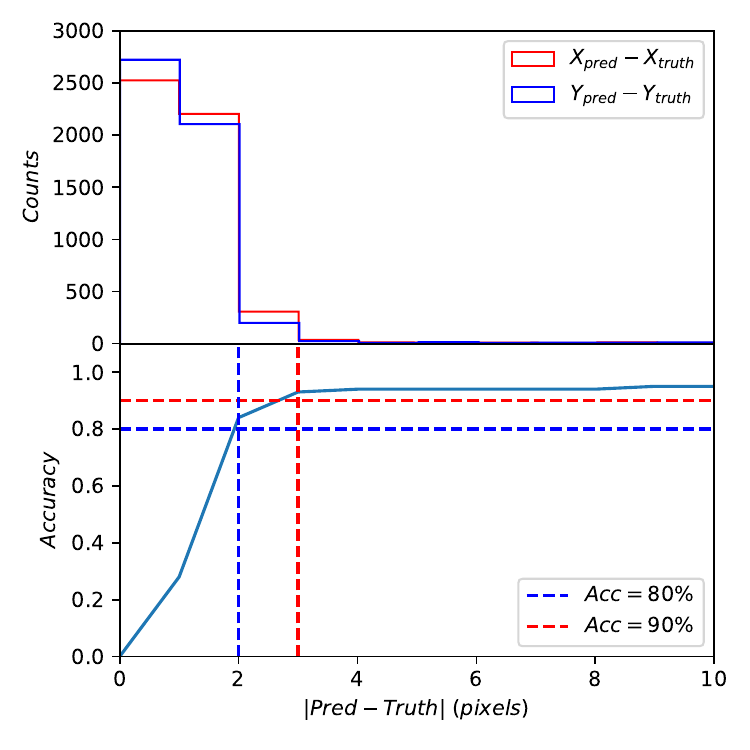}
  \end{tabular}
  \caption{Accuracy of the convolutional ResNet autoencoder described in \Cref{subsection:cnn_autoencoder}. \textit{Left:} $(x, y)$ predicted coordinates of the BCGs as a function of their true $(x, y)$ coordinates as given in the cosmoDC2 simulation. The figure shown here is obtained for the BCGs in the evaluation sample. The color bar represents the number density of points. \textit{Right}: (Top panel) Histogram of the difference between the predicted and true (x, y) coordinates of the BCGs in the evaluation sample as detected by the ResNet autoencoder. The red and blue histograms represent the $x$ and $y$ coordinates respectively. (Bottom panel) Accuracy of the ResNet autoencoder depending on the defined resolution considered for a well-detected BCG. The blue and red lines correspond to an accuracy of 80\% and 90\% respectively.}
    \label{fig:autoencod_predtruth_histdiff_acc}
\end{figure*}

We here identify the BCG position as detected by the convolutional ResNet autoencoder described in \Cref{subsection:cnn_autoencoder} as the pixel on the output probability map with the highest probability.
We first discuss the results obtained by training the neural networks using probability maps with the BCG as a circular gaussian convoluted source. We will then compare these results with those obtained by using the Gaussian convoluted segmentation maps.

Similarly to the results found in the previous \Cref{subsection:perf_resnet} (see \Cref{fig:resnet_predtruth_histdiff_acc} and \Cref{fig:resnet_zdiff}), the coordinates of the detected BCGs are very well correlated with those of their true BCGs (see left panel of \Cref{fig:autoencod_predtruth_histdiff_acc}) with correlation coefficients of $r$ = 0.96 for both $(x, y)$ coordinates. There is also no bias with redshift. 

\begin{figure*}[t]
\centering
\includegraphics[width=\hsize]{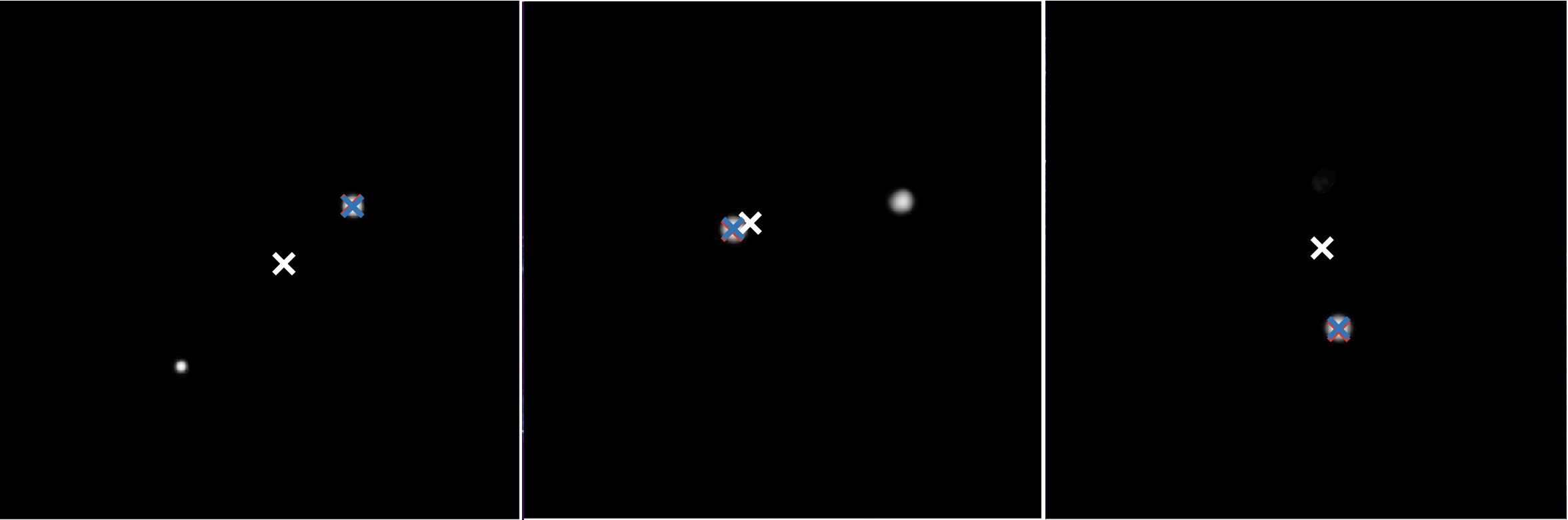}
\caption{Example of the same three clusters as shown in \Cref{fig:resnet_detecthaloex}. The probability maps were obtained using the ResNet autoencoder described in \Cref{subsection:cnn_autoencoder}. The blue cross correspond to the position of the detected BCG, associated with the pixel with the highest probability. For comparison are also overlaid the red and white crosses in \Cref{fig:resnet_detecthaloex} corresponding to the position of the predicted BCG by the convolutional ResNet described in \Cref{subsection:cnn_resnet} and the true BCG respectively.}
\label{fig:autoencod_detecthaloex}
\end{figure*}

The accuracy of the autoencoder compared to the simple ResNet is significantly better by a magnitude. Indeed, we find that the accuracy of the autoencoder increases the accuracy to 95\%, also with much better resolution. As can be seen in the right panel of \Cref{fig:autoencod_predtruth_histdiff_acc}, 93\% of detected BCGs have an offset of less than 3 pixels. This can be explained by the difference in outputs between the convolutional ResNet and the residual autoencoder. On one hand, the simple ResNet compresses all the information contained in the input and outputs a single set of coordinates $(x, y)$. The autoencoder on the other hand restores the information extracted from the input back to its original format. In the case of confusion between several BCG candidates, as could be seen in \Cref{fig:resnet_detecthaloex}, the autoencoder can flag each of these detections with different weights. Such output probabily maps are shown on \Cref{fig:autoencod_detecthaloex}. The clusters shown are the same as the ones shown on \Cref{fig:resnet_detecthaloex}. The true position of the BCG in the cosmoDC2 catalogue is indicated by a red cross, the position of the detected BCG by the ResNet autoencoder associated with the pixel with the highest probability is indicated by a blue cross, and for comparison, the position of the $(x, y)$ coordinates of the detected BCG by the ResNet detailed in \Cref{subsection:cnn_resnet} is indicated by a white cross. In all of these three examples, in which the ResNet described in \Cref{subsection:cnn_resnet} failed to accurately detect the BCG and pointed to the background, the autoencoder ResNet manages to correctly detect them with a probability of more than 90‰. It can be noted that a second source is flagged as a potential BCG on these three examples, although with much lower probability (around 60\% for first two examples, less than 20\% for the third example). The coordinates returned by the ResNet in \Cref{subsection:cnn_resnet} appears to fall in-between the true BCG position and the second flagged source. We confirm in this way that the ResNet in \Cref{subsection:cnn_resnet} appears to have been confused by the presence of several BCG candidates, and the output format did not allow it to properly convey this information.

\begin{figure}
\centering
\includegraphics[width=\hsize]{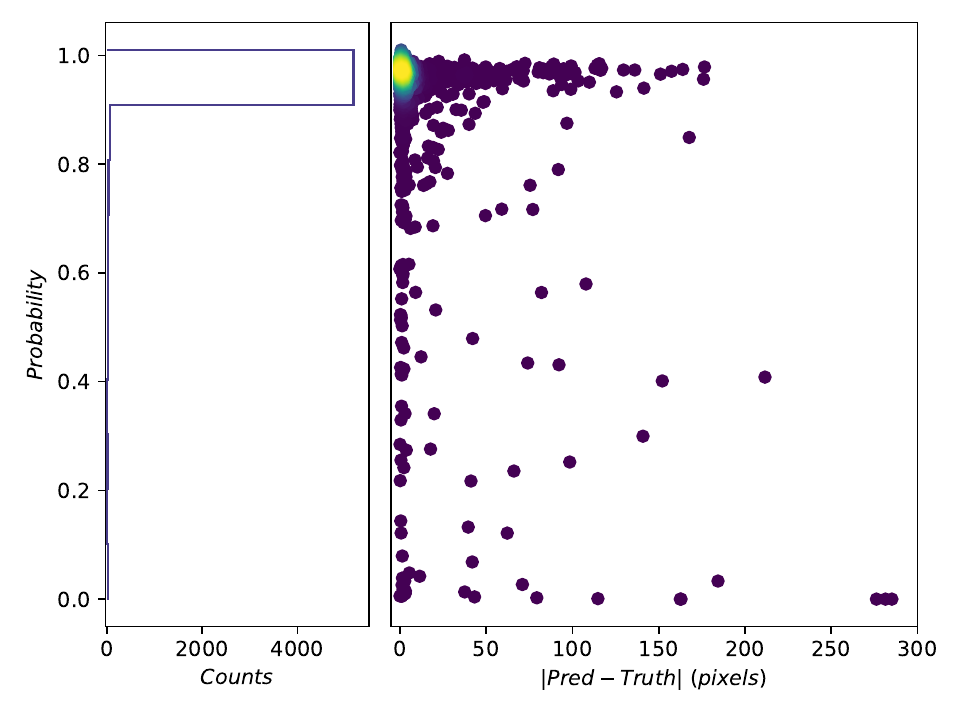}
\caption{\textit{Left panel:} Histogram of the BCG probabilities for the detected BCGs in the evaluation sample. \textit{Right panel:} Detection probabilities of the detected BCGs as a function of the offset between the predicted and true coordinates of the BCG. The colour bar represents the number density of points.}
\label{fig:autoencod_histprobadiff}
\end{figure}

We show in \Cref{fig:autoencod_histprobadiff} the distribution of the probabilities of the detected BCGs in our evaluation sample and the probabilities as a function of the offset between the detected and true coordinates of the BCGs. In total, out of the 5362 clusters in the evaluation sample, 5281 (98\%) of them have a detected BCG with a probability higher than 70\%. However, as illustrated on the right panel of \Cref{fig:autoencod_histprobadiff}, not all of these BCGs are accurate. Indeed, we find that only 5032 (94\%) BCGs out of 5362 have both a high probability and an accurate detected BCG. The difference of 249 galaxies comes mainly from the above mentioned clusters with several BCG candidates. The neural network thus estimated that one of the other candidates was more likely to be the BCG. We note that we do not necessarily consider those as bad detections, as they are very comparable to the BCG defined in the cosmoDC2 catalogue. We also find that about 1\% of our evaluation sample has a well-predicted BCG but a low detection probability of less than 70\%. 
A small cloud of points can be found with probabilities close to zero. Those correspond to clusters for which the neural network could not detect any BCGs. A visual inspection of these clusters reveals that those belong to two main categories: first, BCGs close to another object causing deblending issues; second, red BCGs which have a very low signal-to-noise ratio even on the reddest filter, which makes their detection difficult.

We now discuss the performance of the neural network using more complex segmentation-based probability maps in the training, as opposed to simple circular Gaussian sources. The use of segmentation maps to generate probability maps allows us to retain information on the morphology of the BCG, contrary to the circular sources which just indicate the centre of the BCG. We could assume the use of more complex outputs in the training would result in better performance, as additional information is contained in them. However, unexpectedly, we find that the use of more complex probability maps during the training, based on segmentation maps, does not improve the accuracy of the neural network. The autoencoder performs just as well with both sets of outputs. This may indicate that, on the galactic scale, the morphology of the BCG does not play a significant role in the training, and that the neural network puts more weight on the central luminosity of the galaxy to identify the BCG.

\subsection{Comparison with red sequence based algorithms}
\label{subsection:detection_rs}

In this section, with the aim of justifying the use of deep learning for BCG detection, we compare the efficiency of the neural networks described in this section to more traditional methods which rely only on photometry. In this paper, we compare the results obtained with those of \citet{Chu+21} which is based on a red sequence extraction method. In a first approach, we only consider here a subsample of mid-redshift clusters, from redshifts 0.4 $\leq z \leq$ 0.7. The upper limit at $z \leq$ 0.7 is the usual detection limit of most red sequence-based algorithms, as at higher redshifts, most surveys do not go enough in the infrared to constrain the $4000\AA$ break and thus obtain good photo-$z$ estimations. This study is made considering a subsample of 677 clusters between $0.4 \leq z \leq 0.7$.

\citet{Chu+21} detects the BCG in a sample of 132 clusters observed with the Hubble Space Telescope, from redshift $z = 0.1$ to $z = 1.8$. The method described in that paper was adapted to fit the given data, enabling \citet{Chu+21} to accurately detect all red BCGs in their samples. We can expect the performance of this method to be less efficient for another given set of data, and more tweaking of the parameters would be necessary. Also, the main difference between the sample used in \citet{Chu+21} and this paper is the fact that this paper uses simulation data. We take the same guideline as detailed in \citet{Chu+21} and adapt and try to optimize it for our LSST-DP0.2 simulated data for fair comparison.

Since the cosmoDC2 already provides photometric measurements of galaxies in the simulation, we use their measurements in the retrieved catalogues. Instead of \texttt{SourcExtractor} measured magnitudes, we use the parameters $mag\_\{ugrizy\}\_lsst$ magnitudes. The effective radius and ellipticities are taken as the $size\_true$ and $ellipticity\_true$ parameters respectively. 

We first select all galaxies in the cosmoDC2 simulation in a 1.5 Mpc aperture in the field of the cluster with a magnitude $mag\_r\_lsst <$ 28, the limiting magnitude of the survey. We then filter out foreground galaxies by computing their "pseudo absolute magnitude" which is the absolute magnitude of a given object computed at the cluster's redshift, considering their measured luminosities on the image. Foreground galaxies, via this parameter, would then appear too bright. We thus exclude all objects with pseudo absolute magnitudes brighter than -26 mag in both the $i$ and $z$ bands. Edge-on galaxies are identified if they have ellipticities $ellipticity\_true \leq$ 2.6.

The red sequence is then extracted by selecting all galaxies with a colour $i - z$ that is less than 0.6 mag farther than a model. The model is based on a spectral energy distribution from \citet{bruzual10.1046/j.1365-8711.2003.06897.x}, which assumes a star formation history that decreases exponentially from a single burst at $z_{f}$ = 5 with $\tau$ = 0.5 Gyr, solar metallicity and Chabrier initial mass function.
Sorting through luminosities, we then calculate the galaxy density around every BCG candidate to look for red galaxies overdensities. See \citet{Chu+21} for the detailed procedure.

Doing this, we find that the traditional red sequence algorithm that is used for most wide astronomical surveys, applied to our LSST-DP0.2 simulation, has a 70\% accuracy. The 30\% of bad detections, similar to the two other methods, also come from the presence of multiple BCG candidates. Because of photometric uncertainties, some galaxies, if close in luminosity to the true BCG, may have a slightly brighter measured magnitude which explains its detection instead of the true BCG. Most of the errors, however, come from the bad extraction of the red sequence galaxies with the presence still of interlopers in the red sequence catalogue. These interlopers may remain because of photometric uncertainties (particularly for sources which need to be deblended), or because they may appear to have similar colours to the cluster's galaxies although being foreground sources.
This method also requires more pre-processing with the generation of photometric catalogues and the following red sequence extraction. 
We thus significantly improve the performance of our detection method by more than 20\% by implementing deep learning.

\section{Discussion and conclusions}
\label{section:discussion_conclusion}

In this paper, we make use of machine learning to detect BCGs on optical images, in preparation for the upcoming LSST data release. In that aim, we make use of simulation images from the LSST Data Preview DP0.2 which provides images in all 6 filterbands $ugrizy$, obtained following the simulated LSST cadence up to the 5-year mark. We use two different neural networks: the first one, described in \Cref{subsection:cnn_resnet}, is a convolutional ResNet which takes as input images in all 6 bandpasses and outputs the $(x,y)$ pixel coordinates of the BCG on the images. This very straightforward method returns 80\% of good detections in our evaluation sample. However, we find that this method can not be reliable in cases where the cluster presents multiple BCG candidates with similar properties. In such cases, the ResNet will return a position that is in between the positions of these candidates. The simple binary output format is thus not suited for this study. The lack of a flag to differentiate the good from the bad detections does not enable us to construct a pure and complete BCG catalogue without post-processing. A simple post-processing procedure to check if the coordinates returned by the ResNet are correct could consist, for example, in checking that these coordinates are indeed associated with a bright galaxy in the central region of the image using a source detection code. 

In an effort to bypass this obstacle, we make use of an autoencoder ResNet which is described in \Cref{subsection:cnn_autoencoder}. Contrary to the traditional ResNet which outputs coordinates, the autoencoder outputs a probability map of the same format as the input images. The pixel with the maximum probability is associated with the position of the BCG. The autoencoder has two main advantages as compared to the traditional ResNet: it provides a confidence level, here probabilities, to identify which detections may be faulty; and it indicates the position of all potential candidates with relative weights. The autoencoder increases the performance of the ResNet from 80\% to 95\%. Additionally, we note that in some clusters, the galaxy detected by the autoencoder, which may not coincide with the BCG in the cosmoDC2 catalogue, may be just as good of a BCG candidate. Indeed, some clusters may present not only one but two or more BCGs that differ only very slightly in brightness and size. In occurrence, a supercluster may present one BCG in each of its substructure. For merging clusters, with their BCGs bound to merge together at some point, it is also complicated to determine which is the true BCG. Different definitions of a BCG may lead to the identification of a different BCG. It is still unclear if a slightly brighter but offset massive galaxy would be a better candidate than a slightly less bright but more central galaxy as a BCG. For these reasons, and considering these uncertainties, we could consider that some of the bad detections made by the autoencoder may actually be reasonable. We are therefore likely underestimating the autoencoder's accuracy. The ability of the autoencoder to detect several candidates is also a very strong feature of the network. This feature may be very useful for ICL studies for example, in which the question of whether to consider only one or two BCGs to measure the ICL fraction in clusters is still pending. The autoencoder will allow us not only to identify these clusters with several BCGs, but also to give the positions of these different candidates with relative weights. A simple way to identify these clusters automatically would be to apply a source detection code on the output probability maps and select those with more than one detected sources. It is to be noted however, that the autoencoder described in this paper was not trained with probability maps that contain multiple sources. It may thus be refined by implementing these additional sources in the probability maps used in the training, attributing to these secondary sources a probability that is a function of the ratio of size, luminosity and distance from the true BCG in the cluster.

Our neural networks' main challenge is to differentiate between the different possible candidates in the field. In comparison, algorithms which rely on photometry have to deal with a prior problem which is removing interlopers such as foreground and background sources to extract the red sequence accurately. The presence of a nearby cluster, on top of photometric uncertainties, makes the extraction of the red sequence all the more difficult and these algorithms unreliable for big surveys. The machine learning methods seem to be dealing with these problems better than red sequence finder algorithms.

One common point of contention for all three methods is the deblending of close sources. BCGs hidden in the halo of a foreground source are likely not to be detected by either method. This issue is unfortunately very common in detection problems and is challenging even for neural networks. Efforts to solve this ongoing problem have been made. In particular, \citet{Burke10.1093/mnras/stz2845} have developed new deep learning techniques that have allowed them to cleanly deblend 98\% of galaxies in a crowded field. Their method could be associated with ours in order to boost the performance of our method and potentially retrieve those blended BCGs in our final catalogue.

We can compare the neural networks presented in this paper to one presented by \citet{2025arXiv250200104J}. \citet{2025arXiv250200104J} uses machine learning to detect BCG in large surveys such as the SDSS, using mock observations from The Three Hundred project and real SDSS images for the training. Their network appears to have very good robustness up to $z$ = 0.6 once applied to real data. In comparison, our method seems to be robust up to $z$ = 2.0 but has not been tested on observational data yet. 

It is important to note, once again, that the neural networks described in this paper were trained using simulation data and not real observational data. Observational data differs from simulated data by the presence of noise and artefacts that can not be perfectly reproduced with simulations. Also, the LSST DP0.2 simulation does not include AGNs, strong lenses, solar system objects, or more importantly low surface brightness diffuse features such as tidal streams or ICL which is a very strong feature associated with the stellar halo of the BCG.
As a result, the current trained network can not be applied directly to the future LSST data release. Indeed, studies such as \citet{belfiore2025machinelearninggapreal} for example have shown the gap between real and simulated nebulae and the effect of the nature of the sample on the training of the neural network. The current trained models could be applied on real images by continuing the training using real LSST images, using as training a catalogue of clusters confirmed spectroscopically with identified BCGs. This form of transfer learning from simulation to real observational data would be a better approach than training a new model from scratch using only the future data release images. We can expect the current network to have learned the main features for BCG detection, and so the trained neural weights to be approximately correct. We may hope for the neural networks to learn additional features from the presence of ICL or tidal streams to better detect the BCGs on the real images, and adjust these weights, as those are features which are mostly associated with BCGs. In such a case, we could expect the use of segmentation maps in the training to actually increase the performance of the autoencoder, whereas in this paper, they are not essential.

In the meantime, the performance of the autoencoder applied on the real future data release from LSST may be estimated using data from the HSC-SSP survey. HSC-SSP is very comparable to the future LSST survey in terms of depth, resolution and other imaging properties, but differs mainly by its much smaller footprint and the lack of the $u$-band. The network may be trained again using transfer learning by using images from HSC-SSP to estimate its robustness on real data.

We thus show that machine learning can be a great tool for BCG detection. Although often seen as a black box which lacks physical motivations, the two neural networks presented here perform better than the basic red sequence extractor by a magnitude. The implementation of probability maps also allows us to flag inconclusive detections. 
The resulting BCG catalogue, constructed with the help of machine learning, which will make an inventory of all BCGs in clusters in the southern hemisphere surveyed by LSST over the course of the next ten years, may be used for various studies. 
The combination of studies of the structural physical properties of BCGs, joined with studies of the ICL and tidal streams associated with the BCGs, from local to proto-clusters ($z > 2$), will enlighten our understanding of galaxy formation and evolution in dense environments. The neural networks presented in this paper will enable us to obtain a pure and complete BCG catalogue from the LSST data release in a fast and efficient way.

\begin{acknowledgements}
This research utilized the Sunrise HPC facility supported by the Technical Division at the Department of Physics, Stockholm University. 
This paper makes use of LSST Science Pipelines software developed by the Vera C. Rubin Observatory. We thank the Rubin Observatory for making their code available as free software at \href{https://pipelines.lsst.io}{https://pipelines.lsst.io}.  This work has been enabled by support from the research project grant ‘Understanding the Dynamic Universe’ funded by the Knut and Alice Wallenberg Foundation under Dnr KAW 2018.0067. JJ acknowledges the hospitality of the Aspen Center for Physics, which is supported by National Science Foundation grant PHY-1607611. The participation of JJ at the Aspen Center for Physics was supported by the Simons Foundation. JJ and LD acknowledge support by the Swedish Research Council (VR) under the project 2020-05143 -- ``Deciphering the Dynamics of Cosmic Structure". JJ and LD further acknowledge support by the Simons Collaboration on “Learning the Universe”. This work has been done within the Aquila Consortium
(\href{https://www.aquila-consortium.org}{https://www.aquila-consortium.org}).

\end{acknowledgements}

\bibliographystyle{aa}
\bibliography{mybib}

\end{document}